\documentclass[preprint,showpacs,preprintnumbers,amsmath,amssymb,nofootinbib,natbib]{revtex4}

\usepackage{graphicx}
\usepackage{epsfig}		
\usepackage{dcolumn}
\usepackage{bm}
\def\0{\mbox{\tiny $0$}}
\def\1{\mbox{\tiny $1$}}
\def\2{\mbox{\tiny $2$}}
\def\3{\mbox{\tiny $3$}}
\def\4{\mbox{\tiny $4$}}
\def\5{\mbox{\tiny $5$}}
\def\6{\mbox{\tiny $6$}}
\def\7{\mbox{\tiny $7$}}
\def\8{\mbox{\tiny $8$}}
\def\9{\mbox{\tiny $9$}}

\DeclareMathOperator{\sech}{sech}

\DeclareMathOperator{\csch}{csch}

\begin{document}

\title{Perturbatively deformed defects in P\"oschl-Teller-driven scenarios for quantum mechanics}

\author{Alex E. Bernardini}
\email{alexeb@ufscar.br}
\affiliation{Departamento de F\'{\i}sica, Universidade Federal de S\~ao Carlos, PO Box 676, 13565-905, S\~ao Carlos, SP, Brasil}
\author{Rold\~ao da Rocha}
\email{roldao.rocha@ufabc.edu.br}
\affiliation{CMCC,
Universidade Federal do ABC 09210-580, Santo Andr\'e, SP, Brazil}
\date{\today}

\begin{abstract}
P\"oschl-Teller-driven solutions for quantum mechanical fluctuations are triggered off by single scalar field theories obtained through a systematic perturbative procedure for generating deformed defects.
The analytical properties concerning the quantum fluctuations in one-dimension, zero-mode states, first- and second- excited states, and energy density profiles are all obtained from deformed topological and non-topological structures supported by real scalar fields.
Results are firstly derived from an integrated $\lambda \phi^4$ theory, with corresponding generalizations applied to starting $\lambda \chi^4$ and {\em sine}-Gordon theories.
By focusing our calculations on structures supported by the $\lambda\phi^4$ theory,
the outcome of our study suggests an exact quantitative correspondence to P\"oschl-Teller-driven systems.
Embedded into the perturbative quantum mechanics framework, such a correspondence turns into a helpful tool for computing excited states and continuous mode solutions, as well as their associated energy spectrum, for quantum fluctuations of perturbatively deformed structures.
Perturbative deformations create distinct physical scenarios in the context of exactly solvable quantum systems and may also work as an analytical support for describing novel braneworld universes embedded into a $5$-dimensional gravity bulks.
\end{abstract}

\date{\today}
\maketitle

\section{Introduction}

Defect structures of topological (domain walls) and non-topological (bell-shaped lumps) origin are finite energy static solutions of classical field theories.
Besides their notwithstanding relevance in the construction of nonlinear theories \cite{BookA,BookA22}, defect structures have become very popular in high energy physics, for describing braneworld models \cite{Brane,BraneB,05,06,07,08,09}, phase-transitions and seeds for structure formation of the very early Universe \cite{01,02,03,03B}, or even the existence of magnetic monopoles in the context of the particle physics \cite{010}.
In particular, kink-like structures have been considered either for restoring the symmetry of inflationary universes \cite{012,0011}, or for triggering off mass generation mechanisms for fermions under some sort of symmetry breaking process \cite{017,0018}. 
In solid state physics \cite{Books}, topological defects are namely relevant in describing charge transference in diatomic chains \cite{014,0015,0016}, and also in encompassing {\em skyrmion} models \cite{010B}.
Finally, on the front view of technological applications, high-speed packet-switched optical networks have also included non-topological lump-like models as drivers of bright solitons in optical fibers \cite{017A,0018A}.

Deforming defect strategies for generating novel analytical scenarios of topological (kink-like) and non-topological (lump-like) structures exhibiting some kind of either physical or mathematical appeal have already been largely explored in the literature 
\cite{01,02,03,04,0422,BSG}.
For instance, the sine-Gordon (SG) model \cite{BookB,BookB2,SG01,SG02,SG03,SG04,SG05,rocha11,rocha12,kr,yux} is also known by working as the hedge of an enormous variety of reshaped topological models.
Obtaining analytically manipulable defect structures, in the most of times, solutions of  nonlinear partial/ordinary differential equations, frequently requires some outstanding mathematical technique as a guidance for creating resolvable analytical protocols \cite{04,0422,Bas01,Bazeia,Bas02}.

The perturbative procedure proposed here is possibly much more simplistic than some
well-consolidated strategies \cite{BookA,BookA22,Books,BooksAAA,BooksBBB}, for instance, as those used for obtaining cyclically deformed defect structures \cite{AlexRoldao,Mariana,Mariana2}.
Nevertheless it brings up an advantageous, and eventually unique, connection with P\"oschl-Teller potentials in quantum mechanics \cite{PT}.

The P\"oschl-Teller potential is a particular class of quantum mechanical (QM) potentials for which the one-dimensional Schr\"odinger equation can be analytically solved in terms of Legendre polynomials, $L(u)$, when $u$ is identified with either hyperbolic ($\sim \sech(y)$) or trigonometric ($\sim \sec(y)$) functions.
These potentials are circumstantially required in the analysis of topological solutions
bearing nonlinear equations, as solitary waves throughout Bose-Einstein condensates, or in quantum problems supported by some curved geometrical background \cite{Das,Smith,Jatkar,Witten,Ann00,Ann01,Ann02}.
Given such a geometrical nature, one might suppose the existence of some similarities between P\"oschl-Teller QM potentials and the hyperbolic and trigonometric cyclic chains of deformed topological structures as those obtained from Refs.~\cite{AlexRoldao,Mariana}, as it is indeed noticed.

Throughout this paper a set of analytical properties relative to the QM fluctuations in one-dimension scenarios arises due to topological and non-topological scalar field scenarios obtained from perturbatively deformed defect structures.
Perturbative structures admitting an integrated $\lambda \phi^4$ theory are obtained, and straightforward generalizations for deformed $\lambda \chi^4$ and {\em sine}-Gordon theories are identified.
In this context, the quantitative correspondence with P\"oschl-Teller-driven systems, in the scope of QM perturbations, are essential for computing excited states, continuous mode solutions, and the energy spectrum, for the quantum fluctuations of the mentioned perturbatively deformed structures.
In parallel, such novel perturbative solutions for scalar fields can also be introduced into an analytical scheme for constructing brane models of single real scalar fields coupled to $5$-dimensional gravity warped into $4$-dimensions \cite{Gregory,DeWolfe,Gremm,Erlich,Campos}.

The outline of our work is as follows.
In Section II, the regular procedure for obtaining perturbatively deformed defects is introduced. Defect profiles, associated energy densities, and the corresponding driving potentials supported by primitive well-known one-dimensional defect structures are all identified, and analytical expressions are derived as to show some dependence on a running perturbative parameter, $k$.
In Section III, the approximated P\"oschl-Teller-driven solutions for QM fluctuations related to a deformed $\lambda \phi^4$ theory is quantified.
It has been demonstrated that P\"oschl-Teller eigenstates provide a highly satisfactory approximation for QM modes of perturbatively deformed defects up to $\mathcal{O}(k^4)$, with deviations from the exact expressions for excited states and continuous mode solutions analytically computed as to include corrections of $\mathcal{O}(k^2)$.
Finally, our conclusions are drawn in Section IV, as to point to generalizations of our results to braneworld scenarios driven by a real scalar field coupled to $5$-dimensional gravity.


\section{Perturbatively deformed defects}

Let one considers three primitive defect structures as non-dispersive energy solutions of  nonlinear partial differential equations supported by the following scalar field potentials: a $\lambda \phi^4$ theory, a deformed $\lambda \chi^4$ theory, and a deformed {\em sine}-Gordon theory which effectively works as $\lambda \eta^4$ theory. The fields are given in terms of the one-dimensional coordinate, $y$, with corresponding potentials given by
\begin{eqnarray}
U(\phi) &=& \frac{1}{2}(1- \phi^2)^2,\nonumber \\
U(\chi) &=& 2\chi^2(1- \chi), \nonumber\\
U(\eta) &=& \frac{\eta^2}{2}(1- \eta^2),
\label{topo01}
\end{eqnarray}
for which the coupling constants have been absorbed by scalar field and coordinate re-dimensionalization \cite{BookB,AlexRoldao}.

Reporting about a parametrization in terms of generalized BPS functions \cite{BPS,BPS2}, the nonlinear equations of motion for (static configurations of) the associate scalar field can be turned into first-order equations given in terms of an auxiliary superpotential, $u(\varphi)$, such that one may identify the corresponding potential by
\begin{equation}
U(\varphi) = \frac{1}{2}\left(\frac{du}{d\varphi}\right)^2\qquad\mbox{with}\qquad \frac{du}{d\varphi} \equiv u_{\varphi} = \varphi^{\prime} \equiv \frac{d\varphi}{dy},
\end{equation}
for $\varphi \equiv \phi,\,\chi,\,\eta$.
First-order equations can be evaluated as to give \cite{BookB,AlexRoldao}
\begin{eqnarray}
\phi(y) &=& \tanh(y), \nonumber\\
\chi(y) &=& \sech(y)^2, \nonumber\\
\eta(y) &=& \sech(y),
\label{topo02}
\end{eqnarray}
defects which were extensively discussed in the literature, in the context of defect deformation procedures \cite{Bas01,Bas02,Bazeia}, cyclic deformations \cite{AlexRoldao,Mariana}, and brane and particle physics scenarios \cite{Brane,Ber12,Bertolami}.

The perturbative deforming procedure consists in defining an algebraic expression for derivatives of the analytical indefinite integrals of the primitive defect, $\varphi(y)$, in terms of a perturbation parameter, $k$, to yield
\begin{equation}
\varphi(y) \mapsto \varphi_{(k)}(y)  = \frac{1}{\alpha k}\left(\int ds\,\varphi(s)\bigg{\vert}_{s=\beta(y + k)} - \int ds\,\varphi(s)\bigg{\vert}_{s=\beta(y - k)}\right), 
\end{equation}
with $\alpha$ and $\beta$ arbitrary parameters.
From results given by Eq.~(\ref{topo02}), one immediately notices that the $\lambda \phi^4$ leads to perturbed kink solutions, and
deformed $\lambda \chi^4$ and $\lambda \eta^4$ defects lead to perturbed lump-like structures correspondently given by
\begin{eqnarray}
\phi_{(k)}(y) &=&  \frac{1}{2k}\ln\left[\frac{\cosh(y + k)}{\cosh(y - k)}\right],  \nonumber\\
\chi_{(k)}(y) &=&  \frac{1}{2k}(\tanh(y + k) - \tanh(y-k)),  \nonumber\\
\eta_{(k)}(y) &=&  \frac{1}{k}(\arctan\left[\tanh\left(\frac{1}{2}(y + k)\right)\right] - \arctan\left[\tanh\left(\frac{1}{2}(y-k)\right)\right], 
\label{topo03}
\end{eqnarray}
which are strictly connected to integrated $\lambda \phi^4$, $\lambda \chi^4$ and {\em sine}-Gordon theories, respectively for $\phi_{(k)}(y)$, $\chi_{(k)}(y)$, and $\eta_{(k)}(y)$, and where we have set $\beta = 1$.
One also notices that the above defects are symmetric under a parity operation over $k$, $k \mapsto -k$.
The superpotential derivatives, $u^{(k)}_{\varphi}$ (again for $\varphi \equiv \phi,\,\chi,\,\eta$), in the corresponding BPS scheme are given by 
\begin{eqnarray}
u_\phi^{(k)}= \phi^{\prime}_{(k)}(y) &=&  \frac{1}{2k}(\tanh(y + k) - \tanh(y-k)), \nonumber\\
u_\chi^{(k)}= \chi^{\prime}_{(k)}(y) &=&  \frac{1}{2k}(\sech^2(y + k) - \sech^2(y-k)), \nonumber\\
u_\eta^{(k)}= \eta^{\prime}_{(k)}(y) &=&  -\frac{2}{k}
\frac{\sinh[k]\sinh[y]}{\cosh[2k]+\cosh[2y]}, 
\label{topo03B}
\end{eqnarray}
from which one can compute the associated energy densities, $\rho_{(\varphi)}(y) = \varphi^{\prime}(y)^2$.
The kink-like structure, $\phi^{\prime(k)}(y)$, exhibits a topological charge $Q = 2$, and the lump-like structures are non-topological defects, i. e. $Q =0$, where the charge is currently defined in Refs.~\cite{BookA,BookA22} \footnote{Please, c. f. Eq.~(2.58) on page 33, Ref.~\cite{BookA,BookA22}, and Eq.~(3.6) on page 198, Ref.~\cite{Col85}.},
\begin{equation}
Q=\varphi(y\rightarrow + \infty)-\varphi(y\rightarrow -\infty),\qquad\varphi \equiv \phi,\,\chi,\,\eta.
\end{equation}

\begin{figure}
\begin{center}
\hspace{-1 cm}
\includegraphics[scale=0.5]{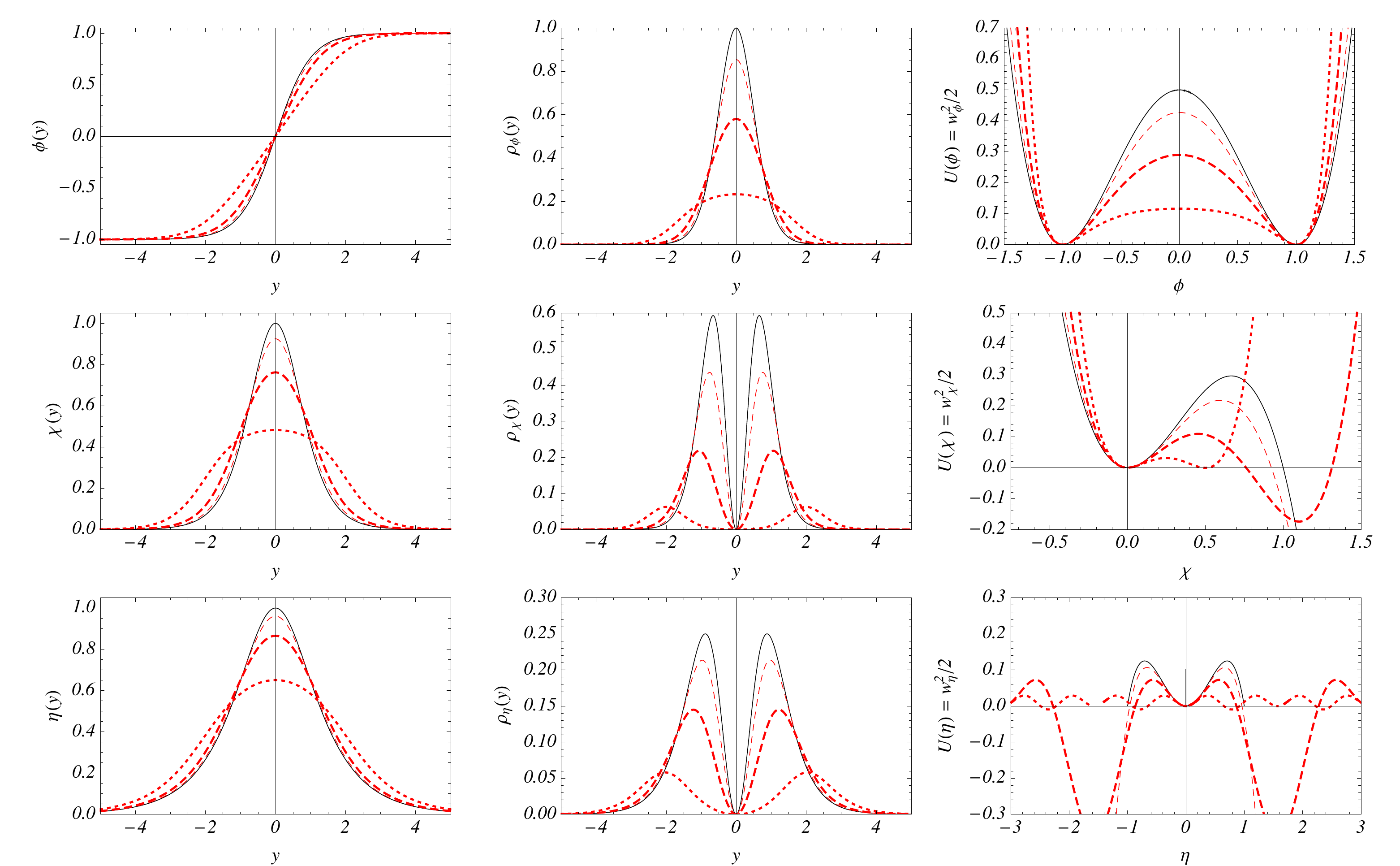}
\end{center}
\caption{(Color online) Perturbatively deformed defect structures obtained from $\lambda \phi^4$ theory, $\phi_{(0)}(y) = \tanh(y)$ (first row), deformed $\lambda \chi^4$ theory, $\chi_{(0)}(y) = \sech(y)^2$ (first row), and deformed $\lambda \eta^4$ theory, $\eta_{(0)}(y) = \sech(y)$ (third row).
Results are for the primitive unperturbed solutions with $k = 0$ (solid black lines), and for the perturbative parameter $k$ with the values of $0.5$ (thin dashed red lines), $1$ (thick dashed red lines), and $2$ (dotted red lines).
The $\lambda \phi^4$ leads to deformed kink solutions.
Deformed $\lambda \phi^4$ and $\lambda \eta^4$ defects lead to lump-like defects.
On the first column one has the defect profile, on the second column one has the energy density profile, and on the third column one has the corresponding scalar field potential.}
\label{Defeitos}
\end{figure}

\begin{figure}
\centering
\includegraphics[scale=0.6]{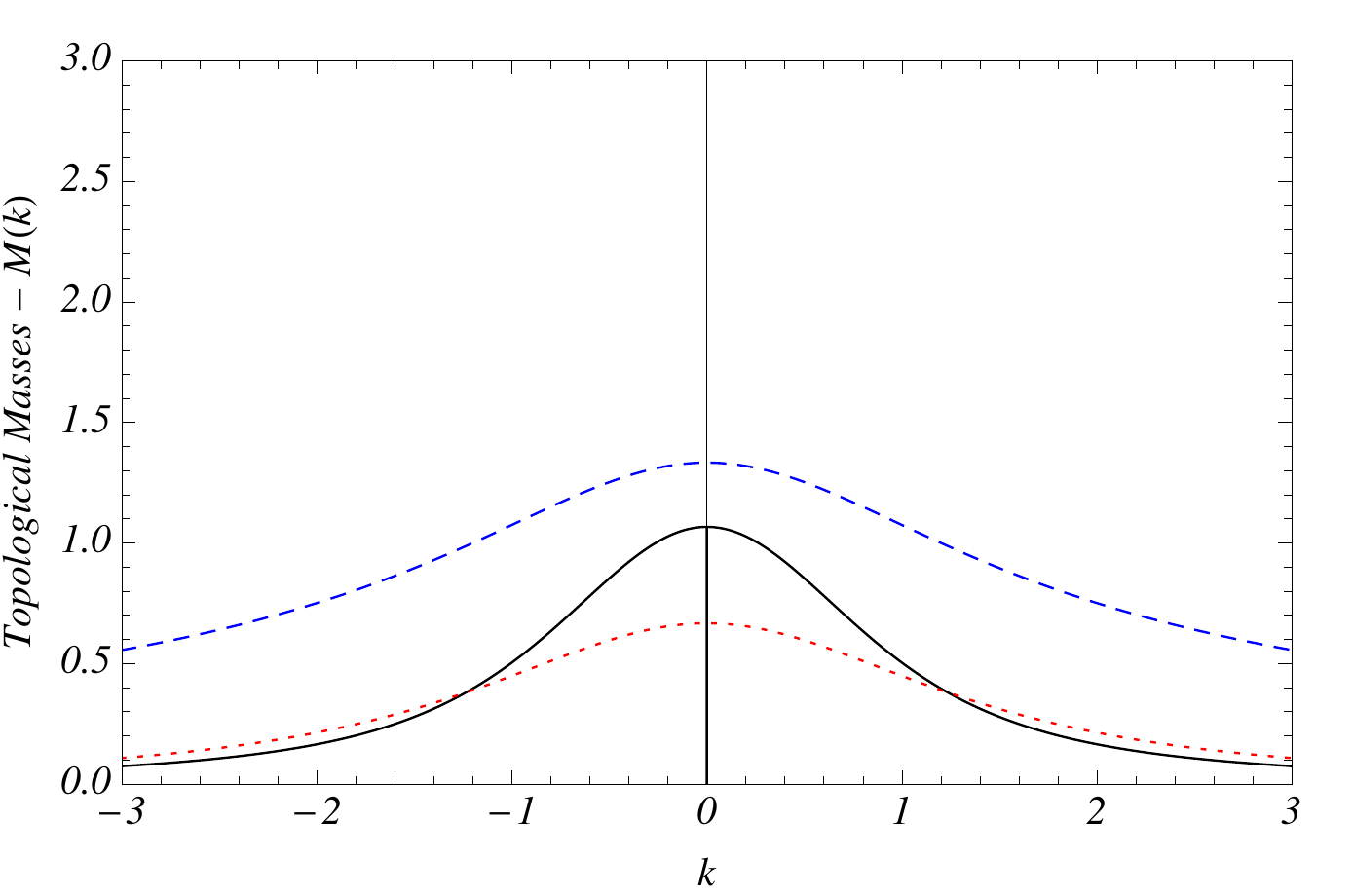}
\caption{(Color online) Topological masses for perturbatively deformed defect structures obtained from $\lambda \phi^4$ theory, $M^\phi_{(k)}$ (blue dashed line), deformed $\lambda \phi^4$ theory, $M^\chi_{(k)}$ (solid black line), and deformed $\lambda \eta^4$ theory, $M^\eta{(k)}$ (dashed red line).}
\label{Massa}
\end{figure}

The defect profiles, the associated energy densities, and the corresponding scalar field potentials are depicted in Fig.~\ref{Defeitos} for the triggering unperturbed solutions and for suitable values of the perturbative parameter, $k$.
By adopting the nomenclature of topological mass, $M$, hereafter as to encompass all definitions of finite energy obtained through the integration of the localized energy density \cite{BookB},
\begin{equation}
M^{\varphi}_{(k)} = 
\left|\int^{+\infty}_{-\infty}dy \,\rho_{(\varphi)}(y)\right|,
\end{equation}
even for those non-topological ($Q =0$) lump-like structures, one obtains
\begin{eqnarray}
M^{\phi}_{(k)} &=&  \frac{2k\coth(2k) - 1}{k^2},  \nonumber\\
M^{\chi}_{(k)} &=&  \frac{\csch(2k)^3(\sinh(6k)+9\sinh(2k)-24k\cosh(2k))}{6k^2},  \nonumber\\
M^{\eta}_{(k)} &=&  \frac{1-k\csch(k)\sech(k)}{k^2}, 
\label{topo04}
\end{eqnarray}
which are depicted in Fig.~\ref{Massa}.

Identifying whether generic solutions, $\varphi(y)$, are stable under tiny time-dependent QM perturbations \cite{Col85} corresponds to the next pertinent issue.
For scalar field potentials encompassing kink- and lump-like defect structures \cite{Bas01,Bas02}, the BPS first-order framework \cite{BPS,BPS2,BookB,Bas01} prescribes perturbative corrections to the field equations of motion in terms of a Schr\"odinger like equation given in terms of field derivatives by
\begin{equation}
\left(- d^2/dy^2 + \varphi^{\prime\prime\prime}/\varphi^{\prime}\right)\psi_{n}(y,t) =\omega_n^2\,\psi_{n}(y,t),
\label{SC222}
\end{equation}
with $\psi_{n}(y,t) = \psi_{n}(y)\exp(-i \omega_n t)$, from which a straightforward manipulation \cite{Col85} leads to the zero-mode wave function (i. e. when $\omega_0 = 0$), given by $\psi_{0}(y,t) \propto \varphi^{\prime}(y)$. The function 
$\varphi^{\prime}(y)$ corresponds to the quantum ground state if $\varphi(y)$ is a kink and, therefore, $\varphi^{\prime}(y)$ has no zeros (nodes).
In this case, there would not be instability for modes with $\omega_n^2 < 0$.
Kinks indeed exhibit a topological monotonic profile for which the derivatives with respect to the coordinate $y$ have no nodes.
On the other hand, lumps exhibit a non-topological non-monotonic behavior for which the derivatives with respect to $y$ have at least one node.
Kinks give rise to stable zero-mode solutions, which are the ground states of the theory, and lumps admit unstable $n$-level excited state solutions of the Schr\"odinger-like equation (\ref{SC222}) \cite{Mariana2}.

\begin{figure}
\hspace{-1 cm}
\includegraphics[scale=0.5]{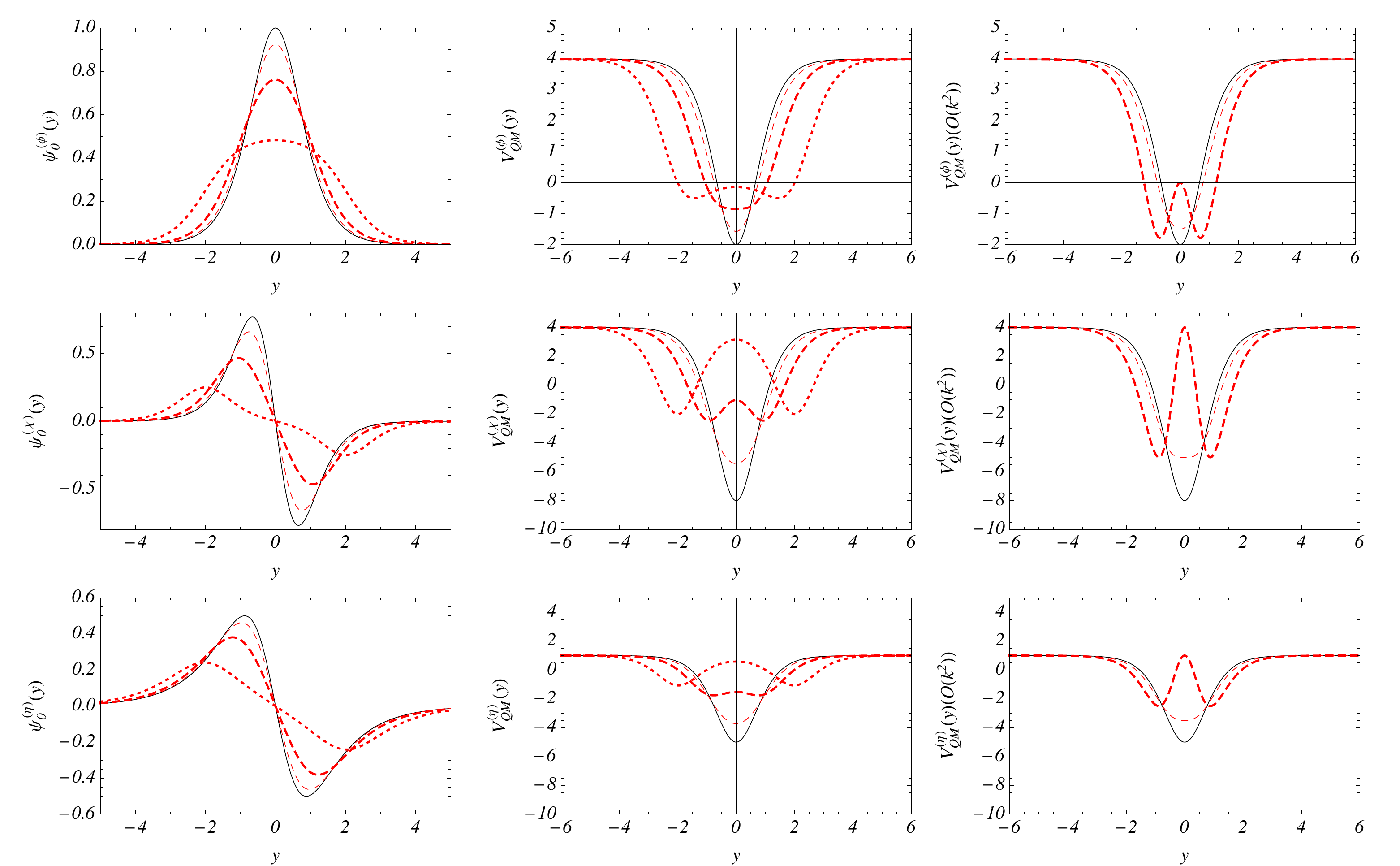}
\caption{(Color online) Normalized zero-mode functions, $\psi_{0}(y)$ (first column), and QM potentials, $V_{QM}(y)$ (second column), for perturbatively deformed defect structures in correspondence with results depicted in Fig.~\ref{Defeitos}, for $\phi$ (first row), $\chi$ (second row) and $\eta$ (third row). 
Again, the results are for the perturbative parameter $k = 0$ (solid black lines), $0.5$ (thin dashed red lines), $1$ (thick dashed red lines), and $2$ (dotted red lines).
The third column depicts the associated P\"oschl-Teller QM potentials with perturbative corrections up to $\mathcal{O}(k^{2})$.}
\label{DefeitosQM}
\end{figure}

By identifying the zero-mode solution with $\psi_{0}(y,t) \propto \varphi^{\prime}(y)$ and the corresponding QM potential with 
$$V^{\varphi}_{QM}(y) = \frac{\varphi^{\prime\prime\prime}(y)}{\varphi^{\prime}(y)}, \qquad \varphi \equiv \phi,\,\chi,\,\eta,$$
one has
\begin{eqnarray}
V^{\phi}_{QM}(y) &=&
2\left(\frac{\sech(y - k)^2 \tanh(y-k)-\sech(y + k)^2 \tanh(y+k)}{\tanh(y + k) - \tanh(y-k)}\right), \nonumber\\
V^{\chi}_{QM}(y) &=& 
4 - 6 \left[\sech(y - k)^2 + \sech(y + k)^2\right], \nonumber \\
V^{\eta}_{QM}(y)  &=& 
\frac{\cosh(4 k) - 8\cosh(2y) - 4\cosh(2k) (2 + 3 \cosh(2y)) +  \cosh(4 y)- 14}{2(\cosh(2k) + \cosh(2y))^2}.
\label{topo06}
\end{eqnarray}
The QM correspondence of the perturbatively deformed defect structures for $\phi_{(k)}(y)$, $\chi_{(k)}(y)$, and $\eta_{(k)}(y)$ are identified in the first and second columns of Fig.~\ref{DefeitosQM}.
In addition, a perturbative expansion of the above obtained QM potentials,
$V^{\varphi}_{QM}(y)$, around $k =0$ leads to approximated P\"oschl-Teller potentials explicitly given by
\begin{eqnarray}
V^{\phi}_{QM}(y) &\approx&
4 - 6\sech(y)^2 - 2 k^2(6 \sech(y)^2 -7 \sech(y)^4)
+\mathcal{O}(k^{\4}), \nonumber\\
V^{\chi}_{QM}(y) &\approx& 
4 - 12\sech(y)^2 - 12 k^2(2 \sech(y)^2 - 3 \sech(y)^4)
+\mathcal{O}(k^{\4}), \nonumber\\
V^{\eta}_{QM}(y)  &\approx& 
1 - 6\sech(y)^2 - 2 k^2(4 \sech(y)^2 -7 \sech(y)^4)
+\mathcal{O}(k^{\4}).
\label{topo06B}
\end{eqnarray}
i. e. $\lim_{k\to 0} V^{\varphi}_{QM}(y) = V^{\varphi}_{PT}(y)$, which are also depicted in Fig.~\ref{DefeitosQM} (third column) up to $\mathcal{O}(k^{\2})$.

In spite of exhibiting a simplistic derivation structure, the above introduced perturbative deforming procedure can be immediately connected to perturbative quantum mechanics, once that a  correspondence with P\"oschl-Teller potentials is noticed.
One may thus obtain the perturbed energy spectrum and the corresponding excited states of the underlying Schr\"odinger quantum model supported by defect structures from (\ref{topo03}) and (\ref{topo06}) just departing from well-known P\"oschl-Teller solutions. These aspects shall be explored in the following section.

\section{Perturbative quantum mechanics supported by the P\"oschl-Teller theory}

\begin{figure}
\centering
\includegraphics[scale=0.65]{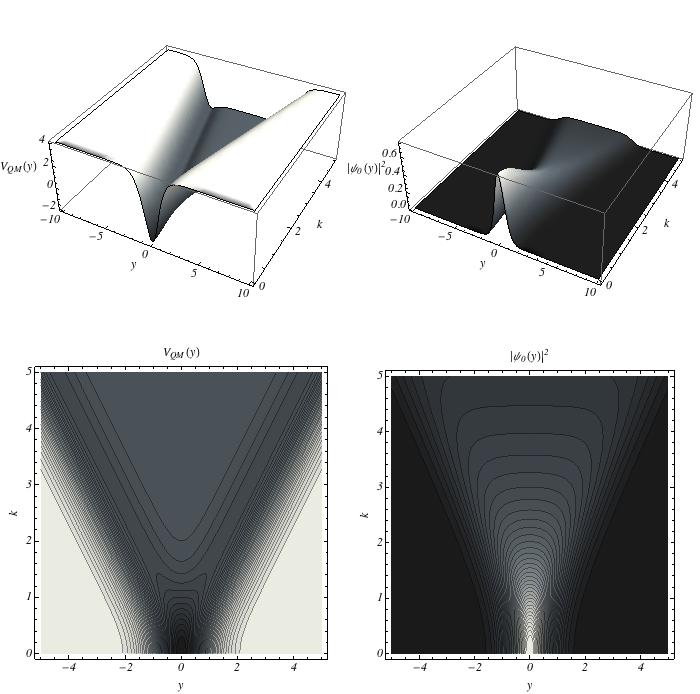}
\caption{QM potential, $V_{QM}(y)$ (first column), and its zero-mode wave function, $\psi_{0}(y)$ (second column), as continuous functions of $k$.
The {\em graytone} scheme has a scale from minimal values (black color) to maximal values (white color) for $V_{QM}(y)$ and $\psi_{0}(y)$.
Increasing $k$ spreads $V_{QM}(y)$ and $\psi_{0}(y)$ along the coordinate $y$.}
\label{WF}
\end{figure}

Let one considers the QM scenario from Eq.~(\ref{SC222}) for perturbatively deformed structures supported by the $\lambda\phi^4$ theory.
Assuming that results for $\phi_{(k)}(y)$, $\chi_{(k)}(y)$, and $\eta_{(k)}(y)$ are qualitatively redundant, hereon only the model supported by $\phi_{(k)}(y)$ shall be considered along the subsequent calculations (and the index $^\phi$ shall be suppressed from the notation).
The overall behavior of the QM potential, $V_{QM}(y)$, and its zero-mode wave function, $\psi_{0}(y)$, as a continuous function of $k$ can be depicted in Fig.~\ref{WF}.
Notice that the original P\"oschl-Teller potential, $V_{QM}(y)$ for $k = 0$, with corresponding zero-mode states, is continuously deformed into a double-well system, as $k$ increases.
If $k < 1$, QM approximated solutions can be perturbatively obtained.

For the P\"oschl-Teller potential obtained from the zero order term in the $k$ expansion from Eq.~(\ref{topo06B}),
\begin{eqnarray}
V_{PT}(y) &=&\lim_{k\to 0} V_{QM}(y) = 
4 - 6\sech(y)^2
\label{topo10}
\end{eqnarray}
one easily identifies the solutions of the corresponding Schr\"odinger equation by
the normalized zero-mode wave function,
\begin{eqnarray}
\psi_{0}(y) = \frac{\sqrt{3}}{2}\sech(y)^2,
\label{topo11}
\end{eqnarray}
the normalized first excited state,
\begin{eqnarray}
\psi_{1}(y) = \frac{\sqrt{3}}{2}\sinh(y)\sech(y)^2,
\label{topo11}
\end{eqnarray}
and the continuous-mode,
\begin{eqnarray}
\psi_{(q)}(y) = \mathcal{N}_{(q)}^{-\frac{1}{2}} \exp(i\,q\,y)(3 \tanh(y)^2 - 1 - q^2 - 3 i\, q \tanh(y)),
\label{topo11}
\end{eqnarray}
with
$$\mathcal{N}_{(q)} = 2 L (4 + 5 q^2 + q^4) -6 \tanh(L) (2 + q^2 - \sech(L)^2),$$ for which  non-vanishing values for the continuous modes just assumed in the interval of $y$ between $-L$ and $+L$.
The state with $q = 0$ can also be misinterpreted as a second excited state.
Concerning the boundary properties, since it is out of the scope of our calculations, the wave number, $q$, is not quantized, and the mode $\psi_{(q)}(y)$ is properly treated as a continuous spectrum wave function\footnote{For $q \to q_n$ quantization procedures one can consult Ref.~\cite{BookB} p. 141.}. 

\begin{figure}
\includegraphics[scale=0.7]{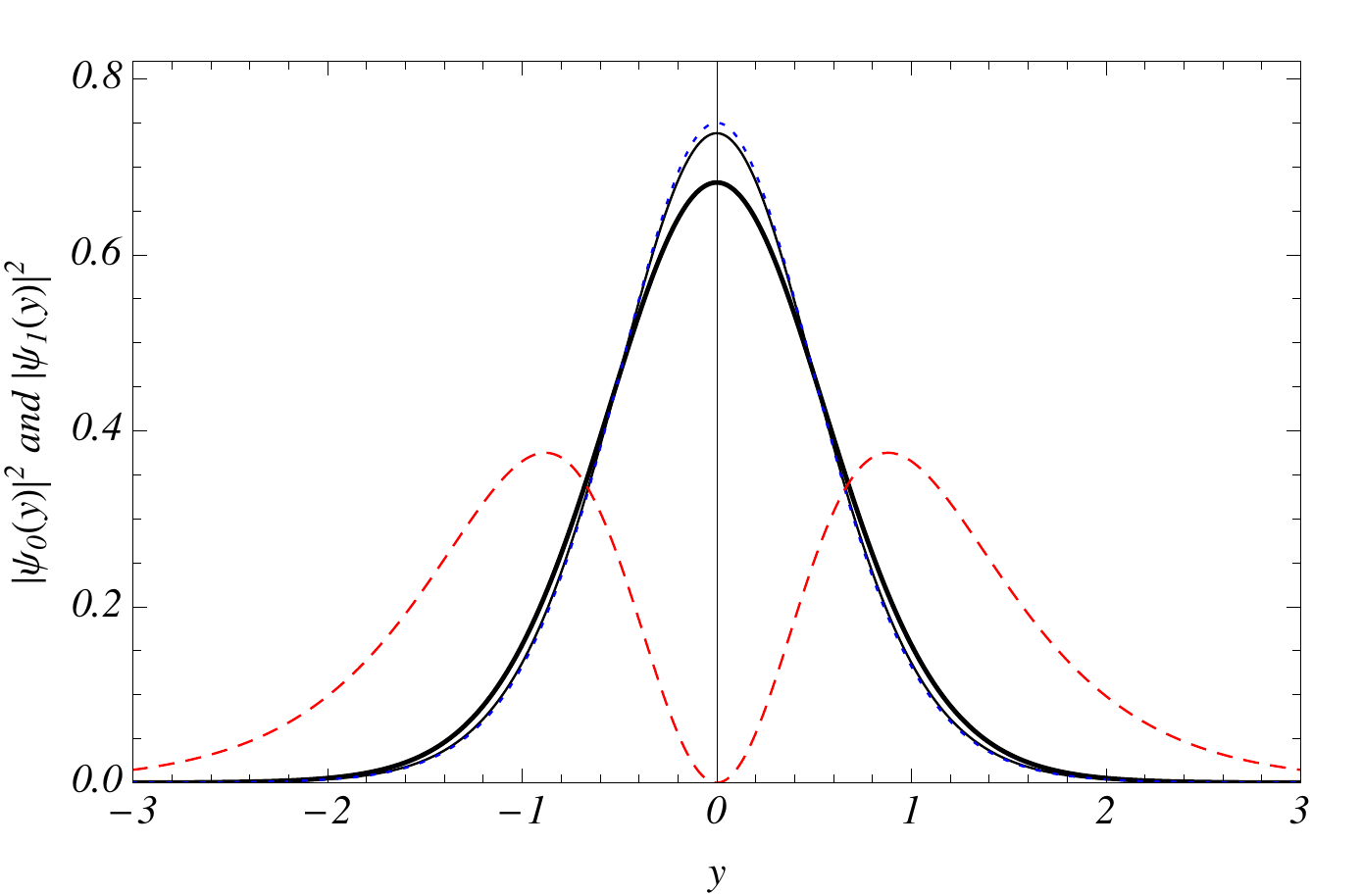}
\caption{(Color online) Normalized squared modulus of exact zero-mode states, $\vert\psi_{0}(y)\vert^{2}$ for $k=0.2$ (thinnest  black line) and $k= 0.5$ (thickest black line) compared with P\"oschl-Teller zero-mode and first excited states, $\vert\psi_{0}(y)\vert^{2}$ (dotted blue line) and $\vert\psi_{1}(y)\vert^{2}$ (dashed red line).}
\label{Wave0e1}
\end{figure}

The discrete level eigenvalues (c. f. Eq.~(\ref{SC222})) are correspondently given by \cite{BookB}
\begin{eqnarray}
\omega_0^2 &=& 0,\\
\omega_1^2 &=& 3,
\label{topo12}
\end{eqnarray}
and, for the continuous mode, by
\begin{eqnarray}
\omega_{(q)}^2 &=& 4 + q^2  \qquad(\to\omega_2^2 = 4).
\label{topo13}
\end{eqnarray}
 By treating $V_{QM}(y)$ under QM perturbations, one has
\begin{eqnarray}
V_{QM}(y) &=& V_{PT}(y) + k^2\delta V(y) + \mathcal{O}(k^4),
\label{topo14}
\end{eqnarray}
with $\delta V(y) = 12 \sech(y)^2 -14 \sech(y)^4$, which provides a satisfactory approximation for the quantum system if one considers the perturbative parameter, $k^2 < 1$. 
By identifying the energy perturbation as
\begin{eqnarray}
\delta\omega_{0,1,2,(q)}^2(k) &=& 
\omega_{0,1,2,(q)}^2(k) - \omega_{0,1,2,(q)}^2(0) = k^2\left(\int_{-b}^{+b} dy\, \vert\psi_{0,1,2,(q)}(y)\vert^2 \delta V(y)\right) + \mathcal{O}(k^4),
\quad\label{topo15}
\end{eqnarray}
with $b = \infty$ for levels $0$, and $1$ and with $b = L$ for continuous levels, for $q\in(-\infty,+\infty)$, one should have
\begin{eqnarray}
\omega_0^2(k) &=& 0 +  \frac{32}{105}k^4,\\
\omega_1^2(k) &=& 3 -  \frac{8}{5}k^2 +  \frac{24}{35}k^4,
\label{topo15B}
\end{eqnarray}
where the corrections have been extended up to $\mathcal{O}(k^4)$, and
\begin{eqnarray}
\omega_2^2(k) &=& 4 + k^2\frac{\sech(L)^7 (360 \sinh(L) - 147 \sinh(3L) + 31 \sinh(5L) - 2 \sinh(7L))}{120 L + 90 (\sech(L)^2) \tanh(L)} + \mathcal{O}(k^4),\nonumber\\
\omega_{(q)}^2(k) &=& 4 + q^2 + 2 k^2 \frac{\tanh(L)}{15} \,F(q,L) + \mathcal{O}(k^4),
\label{topo15B}
\end{eqnarray}
with\footnotesize
$$F(q,L) =
 \frac{ (4 + q^2) (41 + 35 q^2) \sech(L)^2 - 63 (4 + q^2) \sech(L)^4 + 135 \sech(L)^6 - 4 (4 + q^2) (2 + 5 q^2)}{(4 + 5 q^2 + q^4) L - 3 \tanh(L) (2 + q^2 - \sech(L)^2)},$$
\normalsize
which vanishes for the limits of boundary conditions going to infinity, $L \to \infty$.
Fig.~\ref{Wave0e1} shows the normalized squared modulus of exact zero-mode states, $\vert\psi_{0}(y)\vert^{2}$ for $k=0.2$ and $k =0.5$ compared with P\"oschl-Teller zero-mode and first excited states.
Given the orthogonality properties of the wave functions, there is no contribution from perturbative corrections of order $\mathcal{O}(k^2)$, which become relevant for $\mathcal{O}(k^4)$ or higher.
Therefore, the perturbative analysis shows that P\"oschl-Teller eigenstates provides a highly satisfactory approximation for QM modes of perturbatively deformed defects up to $\mathcal{O}(k^4)$

\begin{figure}
\centering
\includegraphics[scale=0.58]{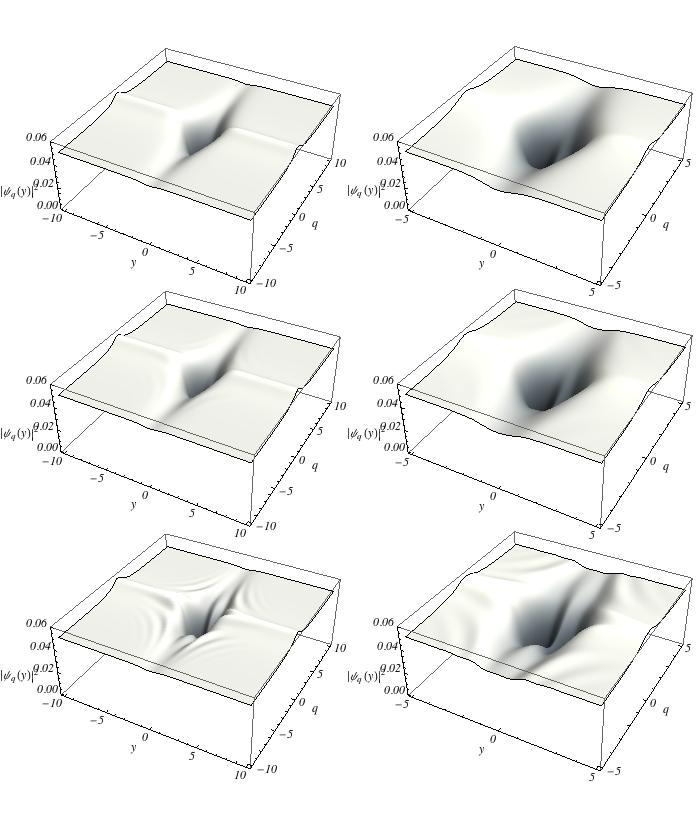}
\caption{Normalized squared modulus of continuous states, $\vert\psi_{(q)}(y)\vert^{2}$ as function of continuous values for the wave vector number, $q$, and coordinate, $y$, assuming that the waves travel from $y = -L$ to $y = L$ with $L = 10$ (first column) and $L = 5$ (second column).
Increasing values of $L$ suppress the perturbation due to the bound states on the continuous modes. For $L$ tending to $\infty$, the perturbative resonances are completely suppressed for larger values of the wave vector, $q$. 
The plots are for $k = 0$ (first row), namely, the unperturbed scenario, and for $k = 0.2$ (second row) and $k = 0.5$ (third row).
The {\em graytone} scheme has a scale from minimal values (black color) to maximal values (white color) of $\vert\psi_{(q)}(y)\vert^{2}$.}
\label{Continuous}
\end{figure}

In the same way, the continuous modes are unperturbed in the limit where $L \to \infty$. Notice from Fig.~\ref{Continuous} that the smaller is the box width, $2L$, the less relevant are the perturbations over the normalized squared modulus of continuous states, $\vert\psi_{(q)}(y)\vert^{2}$ (overall the continuous spectrum of the wave vector number, $q$).
The perturbation due to the bound states, $\psi_{0}$ and $\psi_{1}$, on the continuous modes are suppressed as $L$ increases and tends to $\infty$. It has been evinced for larger values of the wave vector, $q$. For $L$ going to $\infty$, there should be no perturbative resonance imprints on the continuous wave function spectrum.
It is consistent with the observation that $\lim_{L\to\infty} \omega_{2,(q)}^2(k) = \omega_{2,(q)}^2(0)$. 

To sum up, as pointed out in the beginning of this section, the above analysis can be straightforwardly extended to the previously introduced $\chi_{(k)}(y)$ and $\eta_{(k)}(y)$ theories.

\section{Conclusions}

A systematic procedure supported by perturbatively deformed defects has impelled the investigation of approximated P\"oschl-Teller-driven QM systems and braneworld scenarios which encompass a real scalar field coupled to $5$-dimensional gravity.
Analytical profiles driven by input models of $\lambda \phi^4$, $\phi_{(0)}(y) = \tanh(y)$, deformed $\lambda \chi^4$, $\chi_{(0)}(y) = \sech(y)^2$, and deformed $\lambda \eta^4$, $\eta_{(0)}(y) = \sech(y)$ theories have been obtained.
Once they are perturbed by a running parameter $k$, deformed models, respectively named by $\phi_{(k)}(y)$, $\chi_{(k)}(y)$, and $\eta_{(k)}(y)$, are identified and exhibited as a natural connection to integrated $\lambda \phi^4$, $\lambda \chi^4$ and {\em sine}-Gordon theories.

In this context, the existence of nonlinearities in the dynamical equations of motion for scalar field theories supposedly brings up some difficulties in obtaining analytical expressions for QM perturbations and driven braneworld solutions.
The procedure introduced here eliminates some preliminary difficulties and may be extended to more complexified scenarios.

By specializing our calculations to structures supported by the $\lambda\phi^4$ theory, the overall behavior of the QM potential, $V_{QM}(y)$, and its zero-mode wave function, $\psi_{0}(y)$, as well as P\"oschl-Teller-driven QM approximated solutions, were perturbatively obtained.
The mapping of solutions onto P\"oschl-Teller-driven systems, even through a perturbative framework, is highly relevant in the sense that it opens an alternative window for computing excited states and continuous mode wave functions, as well as the associated energy spectrum, for quantum fluctuations triggered off by some topological defect families.

On the other hand, our machinery also allows one to build an analytical description of gravitating structures through topological defects that support thick brane scenarios which exhibit, through there energy density profiles, some perturbatively induced internal structure \cite{Bertolami,Randall,Cvetic01,Folomeev,Rey,Skenderis,Kobayashi,Slatyer}.
In particular, in this context, the formation of a thick brane in $5$-dimensional space-time has been investigated for warped geometries of $AdS5$ type once induced by scalar matter dynamics \cite{Adrianov}.
Given the analytical profiles here obtained, the same perturbatively deformed models can also be helpful in identifying braneworld scenarios, where the deformed potentials give rise to thick brane solutions that exhibit internal structures which reveal the presence of multiple phases in the brane.
In addition, integrable models \cite{Bazeia,Brane,Folomeev,Cvetic01} allows one to interpret the $\phi_{(k)}(y)$ model as an analytical description of gravitating defect structures which admit the inclusion of thick branes \cite{Skenderis,Kobayashi,Slatyer,gauss,German:2013sk,Adrianov} with perturbatively induced smooth internal structures.

The outstanding result obtained from such a novel scenario is concerned with the generation of thick branes with internal structures as a perturbative effect, as it has been shown throughout our analysis.

Finally, an additional note relative to the configurational entropy \cite{Gleiser,Gleiser1} of the perturbatively deformed systems can be raised up in the QM framework. Given that the defects $\phi_{(k)}(y)$, $\chi_{(k)}(y)$, and $\eta_{(k)}(y)$ exhibit an evident symmetry under a parity operation over $k$, $k \mapsto -k$, maximal configurational entropies are straightforwardly obtained for $k=0$.
It means that higher the perturbed solutions correlate with preliminary primitive solutions, closer $k$ is to $0$, and the solutions tend to the most ordered system, which, in certain sense, should be expected.

{\em Acknowledgments - The work of AEB is supported by the Brazilian Agencies FAPESP (grant 2015/05903-4) and CNPq (grant No. 300809/2013-1 and grant No. 440446/2014-7). RdR is grateful to CNPq (grants No. 303027/2012-6,  No. 451682/2015-7 and No. 473326/2013-2), and to FAPESP (grant 2015/10270-0) for partial financial support.}


\begin{thebibliography}{99}

\bibitem{BookA}
G. B. Whitham, {\em Linear and Nonlinear Waves}, (Wiley, New York, 1974).
\bibitem{BookA22}
L. Wilets, {\em Non topological solitons}, (World Scientific, Singapore, 1989).
\bibitem{06}
B. Zwiebach, JHEP {\bf 09}, 028 (2000).
\bibitem{Brane}
D. Bazeia, C. Furtado and A. R. Gomes, JCAP {\bf 0402}, 002 (2004).
\bibitem{05}
A. Sen, Int. J. Mod. Phys. {\bf A20}, 5513 (2005).
\bibitem{07}
J. A. Minahan and B. Zwiebach, JHEP {\bf 09}, 029 (2000).
\bibitem{BraneB}
D. Bazeia and L. Losano, Phys. Rev. {\bf D73}, 025016 (2006).
\bibitem{08}
D. Stojkovic, Phys. Rev. {\bf D67}, 045012 (2003).
\bibitem{09}
M. Giovannini, Phys. Rev. {\bf D75}, 064023 (2007).
\bibitem{01}
L. Gunther and Y. Imry, Phys. Rev. Lett. {\bf 44}, 1225 (1980).
\bibitem{02}
J. -C. Fernandez, J. -M. Gambaudo, S. Gauthier, and G. Reinish, Phys. Rev. Lett. {\bf 46}, 753 (1981).
\bibitem{03}
M. El-Batanouny, S. Burdick, K. M. Martini, and P. Stancioff, Phys. Rev. Lett. {\bf 58}, 2762 (1987).
\bibitem{03B}
A. Liddle and D. H. Lyth, {\em Cosmological Inflation and Large Scale Structure}, (Cambridge Univ. Press, 20).
\bibitem{010}
P. A. M. Dirac, Proc. Roy. Soc. {\bf A133}, 60 (1931).
\bibitem{012}
A. Vilenkin and E. P. S. Shellard, {\em Cosmic Strings and Other Topological Defects}, (Cambridge UP, Cambridge, UK, 1994).
\bibitem{0011}
A. Vilenkin, Phys. Rev. Lett. {\bf 72}, 3137 (1994).
\bibitem{017}
R. Jackiw and C. Rebbi, Phys. Rev. {\bf D13}, 3398 (1976).
\bibitem{0018}
J. Goldstone and F. Wilczek, Phys. Rev. Lett. {\bf 47}, 986 (1981).
\bibitem{Books}
A. S. Davidov, {\em Solitons in molecular systems}, (Kluver, Dordrecht, 1981).
\bibitem{014}
 St. Pnevmatikos, Phys. Rev. Lett. {\bf 60}, 1534 (1988).
\bibitem{0015}
J. Z. Xu and J. N. Huang, Phys. Lett. {\bf A197}, 127 (1995).
\bibitem{0016}
J. Z. Xu and B. Zhou, Phys. Lett. {\bf A210}, 307 (1996).
\bibitem{010B}
X. Z. Yu {\it et al.}, Nature {\bf 465}, 901 (2010).
\bibitem{017A}
G. P. Agrawal, {\em Nonlinear Fiber Optics}, (Academic, San Diego, 1995).
\bibitem{0018A}
H. A. Haus and W. S. Wong, Rev. Mod. Phys. {\bf 68}, 423 (1996).
\bibitem{04}
Yu. S. Kivshar and B. A. Malomed, Rev. Mod. Phys. {\bf 61}, 763 (1989).
\bibitem{BSG}
D. Bazeia, L. Losano, J. M. C. Malbouisson and R. Menezes, Physica {\bf D237}, 937 (2008).
\bibitem{0422}
Yu. S. Kivshar, D. E. Pelinovsky, T. Cretegny, and M. Peyrard, Phys. Rev. Lett. {\bf 80}, 5032 (1997).
\bibitem{BookB}
R. Rajaraman, {\em Solitons and Instantons}, (North-Holland, Amsterdam, 1982);
\bibitem{BookB2}
N. Manton and P. Sutcliffe, {\em Topological Solitons}, (Cambridge UP, Cambridge, UK, 2004).
\bibitem{SG01}
J. Rubinstein, J. Math. Phys. {\bf 11}, 258 (1970).
\bibitem{SG02}
M. J. Ablowitz, D. J. Kaup, A. C. Newell and H. Segul, Phys. Rev. Lett. {\bf 30}, 1262 (1973). \bibitem{SG03}
L. D. Faddeev, L. A. Takhtajan and V. E. Zakharov, Sov. Phys. Dokl. {\bf 19}, 824 (1975).
\bibitem{SG04}
J. Hruby, Nucl. Phys. {\bf B131}, 275 (1977).
\bibitem{SG05}
S. Ferrara, L. Girardello and S. Sciuto, Phys. Lett. {\bf B76}, 303 (1978).
\bibitem{rocha11}
  D.~Bazeia, R.~Menezes and R.~da Rocha,
    Adv.\ High Energy Phys.\  {\bf 2014}, 276729
  (2014).
  \bibitem{rocha12}
  D.~Bazeia, L.~Losano, R.~Menezes and R.~da Rocha,  Eur.\ Phys.\ J.\ C {\bf 73}, 2499  (2013).
  \bibitem{kr} 
  H.~R.~Christiansen, M.~S.~Cunha and M.~O.~Tahim,
  Phys.\ Rev.\ D {\bf 82}, 085023 (2010).
  \bibitem{yux} 
  Y.~X.~Liu, L.~D.~Zhang, L.~J.~Zhang and Y.~S.~Duan,
  Phys.\ Rev.\ D {\bf 78}, 065025 (2008).
\bibitem{Bas01}
D. Bazeia, L. Losano and J. M. C. Malbouisson, Phys. Rev. {\bf D66}, 101701 (2002).
\bibitem{Bazeia}
D. Bazeia, J. Menezes and R. Menezes, Phys. Rev. Lett. {\bf 91}, 241601 (2003).
\bibitem{Bas02}
A. T. Avelar, D. Bazeia, L. Losano and R. Menezes, Eur. Phys. J {\bf C55}, 133 (2008).
\bibitem{BooksAAA}
J. D. Murray, {\em Mathematical Biology}, (Springer-Verlag, Berlin, 1989).
\bibitem{BooksBBB}
D. Walgraef, {\em Spatio-Temporal Pattern Formation}, (Springer-Verlag, New York, 1997).
\bibitem{AlexRoldao}
A. E. Bernardini and R. da Rocha, Adv. High Energy Phys. (AHEP) {\bf 2013}, 304980 (2013).
\bibitem{Mariana}
A. E. Bernardini, M. Chinaglia and R. da Rocha, Eur. Phys. J. Plus {\bf  130}, 97 (2015).
\bibitem{Mariana2}
A. E. Bernardini and M. Chinaglia, Mod. Phys. Lett. {\bf A24}, 1550118 (2015).
\bibitem{PT}
G. P\"oschl and E. Teller, Zeitschrift f\"ur Physik, {\bf 83}, 143 (1933).
\bibitem{Das}
A. Das, {\em Integrable Models} (World Scientific, Singapore, 1989).
\bibitem{Smith}
C. J. Pethick and H. Smith, {\em Bose-Einstein Condensation in Dilute Gases} (Cambridge University Press, 2001).
\bibitem{Jatkar}
D. P. Jatkar, Nucl. Phys. {\bf B 395}, 167 (1993).
\bibitem{Witten}
E. Witten, Phys. Rev. {\bf D44}, 314 (1991).
\bibitem{Ann00}
F. Correa, V. Jakubsky and M. S. Plyushchay, Annals of Physics {\bf 324}, 1078 (2009) 
\bibitem{Ann01}
S.-A. Yahiaoui, S. Hattou and M. Bentaiba, Annals of Physics {\bf 322}, 2733 (2007) 
\bibitem{Ann02}
F. Correa and M. S. Plyushchay, Annals of Physics {\bf 322}, 2493 (2007) 
\bibitem{Gregory}
R. Gregory, V. A. Rubakov and S. M. Sibiryakov, Phys. Rev. Lett. {\bf 84}, 5928 (2000).
\bibitem{DeWolfe}
O. DeWolfe, D. Z. Freedman, S. S. Gubser and A. Karch, Phys. Rev. {\bf D62}, 046008 (2000). 
\bibitem{Gremm}
M. Gremm, Phys. Lett. {\bf B478}, 434 (2000).
\bibitem{Erlich}
C. Cs\'aki, J. Erlich, T.J. Hollowood and Y. Shirman, Nucl. Phys. {\bf B581}, 309 (2000).
\bibitem{Campos}
A. Campos, Phys. Rev. Lett. {\bf 88}, 141602 (2002).
\bibitem{BPS}
E. B. Bogomol'nyi, Sov. J. Nucl. Phys. {\bf 24}, 489 (1976).
\bibitem{BPS2}
M. K. Prasad, C. H. Sommerfied, Phys. Rev. Lett. {\bf 35}, 760 (1975).
\bibitem{Bertolami}
A. E. Bernardini and O. Bertolami, Phys. Lett. {\bf B726}, 512 (2013).
\bibitem{Ber12}
A. E. Bernardini and R. da Rocha, Phys. Lett. {\bf B 717}, 238 (2012).
\bibitem{Col85}
S. Coleman, {\em Aspects of symmetry: selected Erice lectures of Sidney Coleman} (Cambridge University Press, 1985).
\bibitem{Randall}
L. Randall and R. Sundrum, Phys. Rev. Lett. {\bf 83}, 3370 (1999); {\bf 83}, 4690 (1999).
\bibitem{Cvetic01}
M. Cvetic and H. H. Soleng, Phys. Rept. {\bf 282}, 159 (1997).
\bibitem{Folomeev}
V. Dzhunushaliev, V. Folomeev and M. Minamitsuji, Rep. Prog. Phys. {\bf 73}, 066901 (2010).
\bibitem{Rey}
M. Cvetic, S. Griffies and S. Rey, Nucl. Phys. {\bf B381}, 301 (1992).
\bibitem{Skenderis}
K. Skenderis and P.K. Townsend, Phys. Lett. {\bf B468}, 46 (1999).
\bibitem{Kobayashi}
S. Kobayashi, K. Koyama and J. Soda, Phys. Rev. {\bf D65} 064014 (2002).
\bibitem{Slatyer}
T. R. Slatyer and R. R. Volkas, JHEP {\bf 0704}, 062 (2007).
\bibitem{Adrianov}
A. A. Andrianov, V. A. Andrianov and O. O. Novikov, Eur. Phys. J. {\bf C73}, 2675 (2013).
\bibitem{gauss}
M.~Dias, J.~M.~Hoff da Silva and R.~da Rocha, Europhys. Lett. {\bf 110}, 20004 (2015).
\bibitem{German:2013sk} 
G.~German, A.~Herrera--Aguilar, D.~Malagon--Morejon, I.~Quiros and R.~da Rocha, Phys. Rev. {\bf D89}, 026004 (2014);  G.~German, A.~Herrera-Aguilar, D.~Malagon-Morejon, R.~R.~Mora-Luna and R.~da Rocha, JCAP {\bf 1302}, 035 (2013).
\bibitem{Gleiser}
M. Gleiser and N. Stamatopoulos, Phys. Lett. {\bf B713}, 304 (2012).
\bibitem{Gleiser1} R.~A.~C.~Correa, A.~S.~Dutra and M.~Gleiser, Phys.\ Lett.\ B {\bf 737}, 388 (2014). 
\end{thebibliography}
\end{document}